# Mending Cracks in Rutile TiO$_2$ with Electron Beam

Silu Guo[1], Hwanhui Yun[1,2], Sreejith Nair[1], Bharat Jalan[1] and K. Andre Mkhoyan[1]*

[1]*Chemical Engineering and Materials Science, University of Minnesota, Twin Cities, USA*
[2]*Korea Research Institute of Chemical Technology, Daejeon, Korea*

*Abstract*

Restructuring of rutile TiO$_2$ under electron beam irradiation driven by radiolysis was observed and analyzed using a combination of atomic-resolution imaging and electron energy loss spectroscopy (EELS) in scanning transmission electron microscopy (STEM). It was determined that a high-energy (80-300 keV) electron beam at high doses ($\gtrsim 10^7$ e/nm$^2$) can constructively restructure rutile TiO$_2$ with an efficiency of $6\times10^{-6}$. These observations were realized using rutile TiO$_2$ samples with atomically sharp nanometer-wide cracks. Based on atomic-resolution STEM imaging and quantitative EELS analysis, we propose a "2-step rolling" model of the octahedral building blocks of the crystal to account for observed radiolysis-driven atomic migration.

*Corresponding author
Email: mkhoyan@umn.edu

*Introduction:*

When energetic electrons interact with crystals, a combination of elastic and inelastic interactions take place. Some inelastic interactions result in displacement of the atoms from their original crystal lattice sites. These inelastic interactions often ultimately limit the precision of measurements with electron probes. Transmission electron microscopes are strongly limited by beam damage, as they commonly use high-energy (60 to 300 keV) electron beams to study atomic structure and electronic properties of crystalline materials [1]. For an atom to be displaced from its lattice site under electron beam irradiation, some (or all) of its chemical bonds must be broken. The mechanisms by which these bonds are broken are either by direct transfer of energy and momentum from incident electrons to the atoms, or ionization of those bonding electrons, or by radiolysis [2,3]. When the energy of electrons in the incident beam is considerably high (~200-300 keV) and crystal consists of light elements ($Z \lesssim 20$) with low binding energies ($\lesssim$ 1-2 eV/bond), the knock-on mechanism dominates, drilling holes in samples [4,5]. In all other cases, the radiolytic mechanism predominates, often amorphizing crystalline samples [2,3,6]. For some crystals and for specific beam energies both mechanisms can be in play with similar probabilities, but those cases are rare. Radiolysis requires formation of electron-hole pairs (excitons) with sufficiently long lifetime ($\gtrsim$ 1 ps) and high energy ($\gtrsim$ 2-3 eV). After recombination, these excitons provide the energy and momentum needed to break the bonds and separate atoms from each other before they can recombine [2]. Radiolytic bond breakage and associated crystal amorphization is well documented in halides [7,8], silicates [9,10], zeolites [11-13], and, recently, in MOFs [14,15]. There are distinct differences between structural changes in a material due to radiolysis and those due to sample heating, which occur under specific conditions [1].

Since the atomistic mechanisms of radiolysis-driven crystal-to-amorphous transformations of materials are not well understood, this study sheds new light on radiolysis-driven atomistic processes by directly observing a surprising process: not destructive amorphization but rather constructive crystal formation. We explored several wide band gap oxides ($SiO_2$, $Al_2O_3$, $TiO_2$, $GeO_2$, $SnO_2$, etc.) that host high-energy excitons ($\gtrsim$ 2 eV) and exhibit crystal structures containing tetrahedral or octahedral building blocks, as the movements of these building blocks can be easier to track compared to those of individual atoms. While many of these oxides meet the requirements for radiolysis under electron beam irradiation, i.e., formation of long lifetime and high energy excitons, we focused on rutile TiO$_2$. This material has a tetragonal crystal structure with space group P4$_2$/mnm (a = 4.594 Å, c = 2.959 Å) [16,17], where each Ti atom is surrounded by six O atoms forming a distorted TiO$_6$ octahedral basic building block. Two neighboring octahedra share an edge at the base with two associated oxygens, and each unit is joined with another two through corner oxygen. Since neighboring octahedra only share edges and not faces, they might be able to rotate or move into the available open space in the structure when bonds between units are broken. Earlier reports documented that rutile TiO$_2$ is indeed susceptible to structural modifications under electron beam irradiation [18,19].

In this letter, we report the results of a STEM study of radiolysis in rutile TiO$_2$. The unique geometry of the sample has allowed us to directly visualize and quantitatively analyze radiolytic changes occurring in the material as a function of electron beam exposure. Analysis of atomic-resolution high-angle annular dark-field STEM (HAADF-STEM) images and EELS data shows that radiolysis can be a constructive force restructuring cracks in rutile TiO$_2$ through relocation of building units. These observations point to the constructive potential of radiolysis, which can help to improve the current techniques of crystal growth.

*Results and Discussion*

The study is conducted on single crystal rutile TiO$_2$ with nanometer-wide cracks. Samples were prepared by growing a 26 nm IrO$_2$ film on a rutile TiO$_2$ substrate using solid-source metal-organic molecular beam epitaxy (SSMO-MBE) [20]. When rutile IrO$_2$ thin film is grown on a rutile TiO$_2$ substrate using SSMO-MBE, epitaxial stresses imposed by the film on the substrate ($\varepsilon$ = 2.2% in ⟨110⟩ direction) induce atomically sharp cracks that span micrometers along the in-plane ⟨110⟩ crystallographic directions and propagate well into the TiO$_2$ substrate along the *c*-axis. Such crack formation in rutile TiO$_2$ due to an anisotropic epitaxial strain is well documented [21,22]. Each crack starts with a single-atom width at its tip and widens to 3 nm at the interface of the TiO$_2$ substrate and IrO$_2$ film, leading to wedge angles of 2-3º (see Fig. S1 in SI). These cracks and their crystallographic orientations are readily identifiable in top-view scanning electron microscopy (SEM) images (Fig. 1(a)). To image the full depth of the crack, cross-sectional samples were prepared using focused ion beam (FIB), which cuts perpendicularly to the cracks. Fig. 1(b) shows low-magnification HAADF-STEM image of one such crack cross-section along with a magnified atomic-resolution image of the neighboring rutile TiO$_2$ crystalline structure. Direct comparison with simulated HAADF-STEM images suggests that these FIB-prepared samples are approximately 50 nm thick (see Fig. S2 in SI).

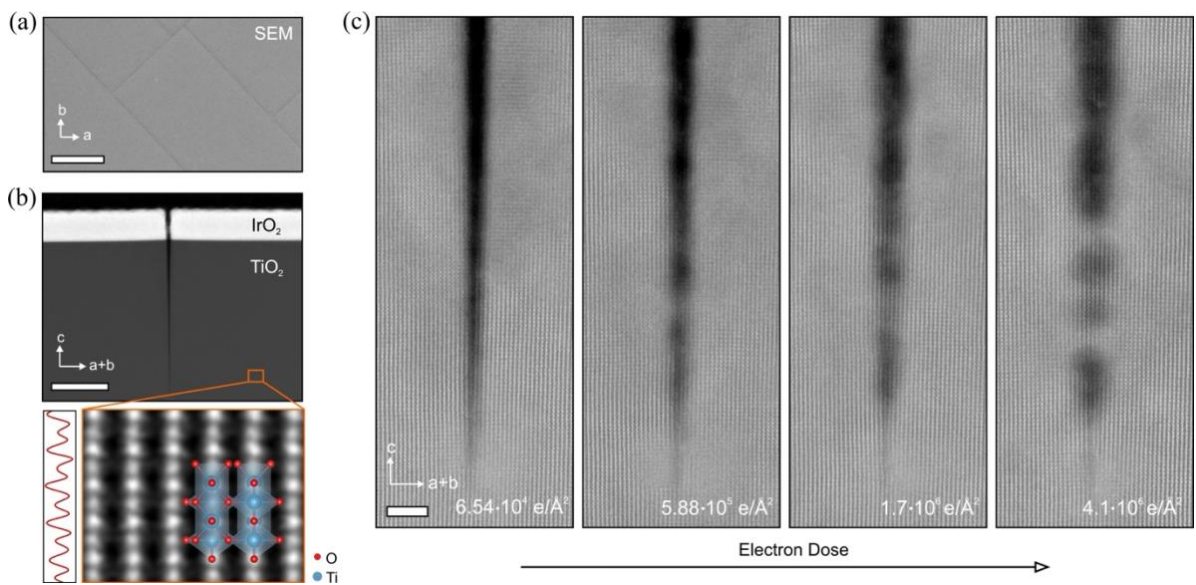

**Fig. 1.** (a) SEM top-view image of a rutile IrO$_2$/TiO$_2$ sample, where the black lines running along the ⟨110⟩ crystallographic directions are the cracks. Scale bar is 1 μm. (b) HAADF-STEM image of a crack propagating through IrO$_2$/TiO$_2$ sample viewed along the [1$\bar{1}$0] direction. Atomic-resolution image of the crystal structure of the rutile TiO$_2$ in this projection shows the arrangements of two distinct atomic columns: columns of only Ti atoms ("dim") and columns of combined Ti and O atoms ("bright"). The line-scan through these columns is shown on the left. Atomic model of the rutile TiO$_2$ is overlayed on the image. (Ti in blue, O in red). Scale bar is 50 nm. (c) A set of HAADF-STEM images showing formation of TiO$_2$ crystal in the crack bridging two sides with increase of electron doses. Scale bar is 3 nm.

To evaluate the effects of electron beam exposure, high-resolution HAADF-STEM time-lapse images of the cracks were obtained. STEM was operated at a beam energy of 200 keV in all these measurements. A set of images obtained from a crack with a wedge angle of 3.6° at a dose rate of 812 e Å$^{-2}$ s$^{-1}$ is shown in Fig.1(c) (see also Video S1 in SI). As can be seen in Fig. 1(c), the crack is self-healing with some areas almost completely filled and others only partially. At the beginning of the beam exposure, at doses $D \lesssim 6.5 \times 10^4$ e/Å$^2$, the crack is still atomically sharp. Starting from an electron dose of approximately $5.9 \times 10^5$ e/Å$^2$, formation of some crystalline structures at several locations along the crack edge are evident. As cumulative electron dose increases, the crack progressively fills.

To further verify the crystalline structure and the composition of the newly formed material in the crack, high-magnification STEM images and EDX elemental maps were obtained from that region. An atomic-resolution HAADF-STEM image of the filled crack after being exposed to an electron dose of $4.1\times10^6$ e/Å$^2$ is shown in Fig. 2(a). The newly formed structure inside the crack is commensurate with the rutile TiO$_2$ outside the crack (Fig. 1(b)). This arrangement of the atomic columns is unique to rutile TiO$_2$ in $\langle 110 \rangle$ orientation. Some fluctuations in HAADF image contrast, observed in the newly formed crystal inside the crack, can be attributed to incompletion of the filling process and to some variations in local thickness of this region (see Fig. S2 in SI). Complementary STEM-EDX maps obtained from the same region confirm that the chemical composition of the newly formed crystal is indeed the same as that of rutile TiO$_2$ (Fig. 2(b)).

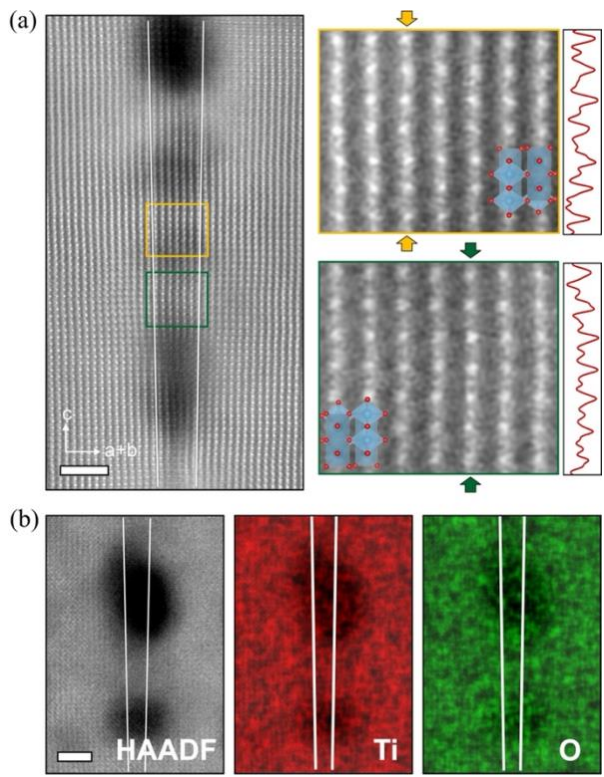

**Fig. 2.** (a) Atomic-resolution HAADF-STEM image of the crack region after $4.1\times10^6$ e/Å$^2$ electron dose exposure. Scale bar is 2 nm. Two magnified regions (highlighted with yellow and green rectangles) show formation of rutile TiO$_2$ crystal inside the crack viewed in [1$\bar{1}$0] orientation. HAADF intensity profiles shown on the right are obtained from two rows of atomic columns indicated by arrows. Atomic model of rutile TiO$_2$ is overlayed on the images. (b) HAADF-STEM image with corresponding STEM-EDX elemental maps from the crack and surrounding regions. Scale bar is 5 nm.

A closer look at the initial restructuring stages reveals interesting details about bridging locations. At an electron dose of $1.5 \times 10^6$ e/Å$^2$, bridging from two opposite sides can be observed at two different locations (Fig. 3). The first bridging location is at the distance $d_1$ = 15 nm from the tip of the crack and the second one is at the same distance from the first: $d_2 = 2d_1$ = 30 nm (Fig. 3 and Fig. S3 in SI). The gaps at these two locations are estimated to be 6.5 Å and 13.1 Å, respectively. These two gap widths are equal to an integral multiple of the unit-cell-length of rutile TiO$_2$ in this projection satisfying the relationship: $d_n \cdot \tan(\alpha) = n \cdot a_{[110]}$, where $n$ = 1, 2, 3, …; $\alpha$ = 2.5° is crack wedge angle; and $a_{[110]}$ = 6.48 Å is the unit cell width of rutile TiO$_2$ along ⟨110⟩ directions. This observation suggests that at these locations, where integer number of rutile TiO$_2$ unit cell lengths can fit, are preferable sites for initial bridging with minimal strain or defects. In between these bridging sites, some crystalline formations can also be seen, particularly at distances of $d_m \cdot \tan(\alpha) = m \cdot a_{[110]}$, where $m = \frac{1}{2}, \frac{3}{2}, \frac{5}{2}, \ldots$ (Fig. 3). But they are not well ordered as the widths of the gaps at these locations are fractional multiples of the unit-cell-length and thus, a perfect rutile structure cannot be satisfactorily accomplished. Additionally, it is observed that the main bridging sites, at distances $d_n$, are surrounded with widened cracks from both sides (Fig. 3), showing that material migrates from neighboring free surfaces into the bridging sites.

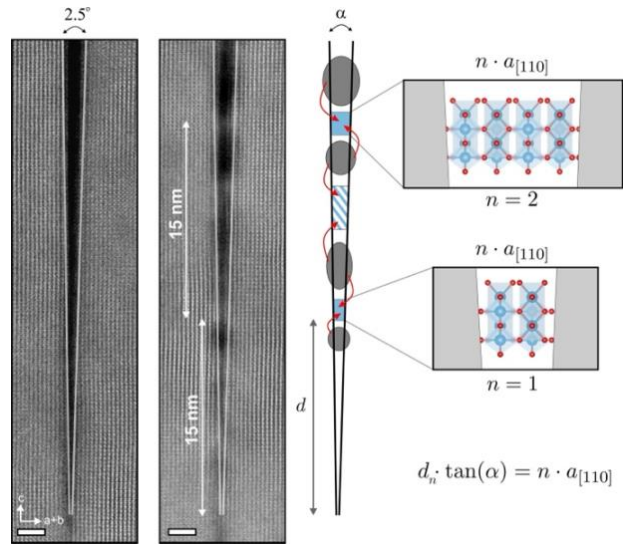

**Fig. 3.** HAADF-STEM images of a crack before and after exposure to the STEM electron beam. Scale bar is 2 nm. The second image is taken at an electron dose of $1.5 \times 10^6$ e/Å$^2$. The model on the right describes

observed initial bridging of the crack at locations where the gap is an integer number of unit cells of rutile $TiO_2$.

This restructuring requires extensive breaking and re-forming of Ti-O bonds near the crack. The release of Ti and O atoms enables them to move out and fill in the crack. Core-level EELS analysis was performed to understand the changes in Ti-O bonds near the crack. Core-loss EELS is particularly sensitive to changes in immediate bonding environment of the atoms in a crystal [23]. Monochromated EELS Ti-$L_{2,3}$-edge fine structure was measured from the crack region as a function of electron dose. EELS spectra were recorded in parallel with HAADF-STEM images (Fig. 4(a)). As can be seen from these HAADF-STEM images, the crack is gradually bridged as the beam dose increases, consistent with our earlier observations. The fine structure of corresponding Ti-$L_{2,3}$ edges, on the other hand, changes non-monotonically. The relative intensities of peaks "b" and "c" depart significantly from the initial fine structure before eventually reverting back. High sensitivity of these two specific peaks in Ti-$L_{2,3}$-edge fine structure to changes in Ti-O bonding state is well understood. In a bulk rutile $TiO_2$ crystal, the EELS Ti-$L_{2,3}$-edge exhibits two groups of peaks, $L_3$ and $L_2$, associated with electronic transitions from initial $2p_{3/2}$ and $2p_{1/2}$ core states to final $3d$ states in the conduction band [24]. In addition, both $L_3$ and $L_2$ edges are divided into $t_{2g}$ and $e_g$ sub-bands due to octahedral crystal field splitting [25-28]. Furthermore, in the $L_3$ edge, the $e_g$ peak splits into peaks "b" and "c" due to Jahn-Teller distortion of the $TiO_6$ octahedra (with $D_{2h}$ symmetry) [26,29]. As a result, these two peaks are highly sensitive to any distortions or bond modifications in $TiO_6$ octahedra. When a Ti-O bond is broken in rutile $TiO_2$ and the O atom is left behind, the original $TiO_6$ reduces to a $TiO_5$ "octahedron" and the $Ti^{4+}$ turns into $Ti^{3+}$. This manifests in the changes in intensities and locations of peaks "b" and "c" [26,30]. By analyzing these two peaks, the changes in Ti valence-state and in Ti-O bonds are evaluated.

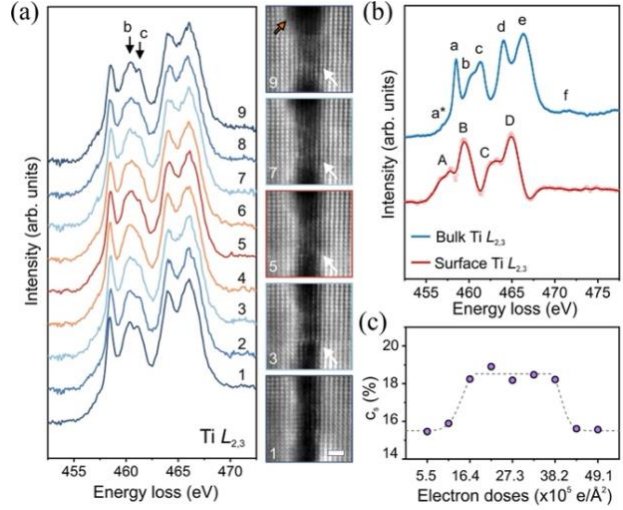

**Fig. 4.** (a) A set of EELS Ti-$L_{2,3}$-edges as function of electron beam doses from a crack region in rutile $TiO_2$. HAADF-STEM images recorded in parallel with EELS measurements are presented on the right showing bridging. Scale bar is 1 nm. The range of electron doses in this measurement is from $5.5 \times 10^5$ e/Å$^2$ (spectrum 1) to $49.1 \times 10^5$ e/Å$^2$ (spectrum 9). The peaks "b" and "c" with most changes are at 460.3 and 461.1 eV, correspondingly. (b) The spectra of bulk and surface Ti $L_{2,3}$-edges with all identifiable features of their fine structure labeled as a*-g and A-E, correspondingly. (c) Concentrations of surface Ti atoms in the exposed crack area ($c_s$) as a function of electron dose determined from Ti $L_{2,3}$-edge spectra in (a) using bulk and surface references in (b).

Since the $TiO_5$ "octahedra" with $Ti^{3+}$ are the dominant units on the surfaces of rutile $TiO_2$, first the characteristic bulk and surface Ti $L_{2,3}$-edges were obtained (Fig. 4(b)). They were deduced from two EELS spectra measured; one from the edge region of the crack and one from a bulk region between two cracks (for more details, see Figs. S4-S5 in SI). As can be seen from Fig. 4(b), they are considerably different from each other due to (4+) and (3+) valence states of Ti in the bulk and on the surface, respectively. The fine structures of these EELS edges are consistent with those from bulk and surface Ti-$L_{2,3}$-edges reported in the literature [26,30]. Using these bulk and surface EELS Ti-$L_{2,3}$-edges as references, each spectrum in Fig. 4(a) was decomposed into bulk and surface components fitted as a linear superposition of two: $I = x \cdot I_s + (1 - x) \cdot I_b$, where $x$ and (1 - $x$) are the relative fractions (or concentrations) of surface and bulk components (for more details, see Fig. S6 in SI). The results are presented in Fig. 4(c). The concentration of surface states initially increases up to 19% from the initial concentration, then recovers back to initial levels after accumulation of electron dose, indicating structural self-healing. When $TiO_x$ octahedra are moving from both sides of the crack into the gap, many Ti-O bonds are breaking, resulting in more surface-like (or $Ti^{3+}$) Ti atoms. Then, when they link to each other from opposite sides, $TiO_6$ octahedra are restored, and the region turns into a crystal with an original rutile structure. When the amounts of Ti and O atoms in this exposed area were evaluated using integrated intensities of Ti $L_{2,3}$- and O $K$-edges, no detectable reduction of either Ti or O was observed (see Fig. S7 in SI), suggesting no change in composition and further supporting radiolysis-dominant filling of the gap in the cracks.

Taking all the observations discussed above into consideration, we propose the following mechanism for structural changes that take place at a crack in rutile $TiO_2$ under electron beam irradiation: sequences of "2-step rolling" bring $TiO_6$ octahedral units from both crack edges into free space to eventually bridge and fill the crack. This "2-step rolling" is illustrated in Fig. 5 (see also SI Video S2 and Fig. S8). As illustrated in Fig. 5, with radiolytic bond breakage, the $TiO_6$ octahedral units from corners, edges, and surfaces can roll and occupy new sites, thus producing net mass transfer from crack edges into the crack gap. Such rolling of the $TiO_6$ octahedra can also take place vertically, parallel to the edges of the crack (see Fig. S8 in SI), which accounts for the relocation of material into the gap bridging sites from nearby areas (Figs. 1, 3 and 4(a)). The formation of many new surface $TiO_x$ octahedra are captured in HAADF-STEM images and Ti-$L_{2,3}$-edge spectra (Figs. 3 and 4). When these new octahedral units meet in the middle of the crack, they unite through corner-sharing oxygens and form the rutile $TiO_2$ crystal structure.

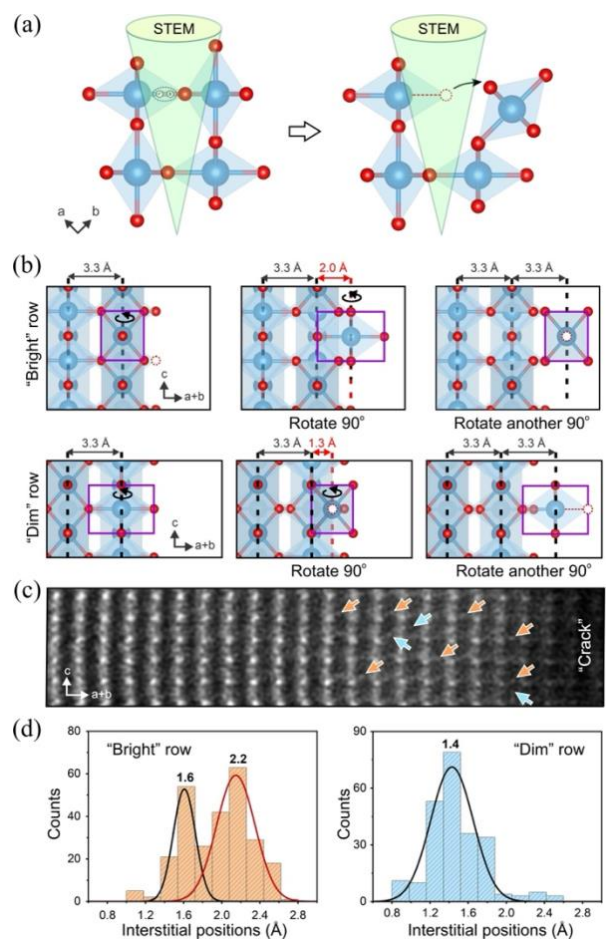

**Fig. 5.** (a) A schematic illustrating radiolysis-driven Ti-O bond breaking between two linking $TiO_6$ octahedra and their separation with aid of exciton ($e^-$-$h^+$ pair) generation. (b) Illustrations of the "2-step" rolling model showing how an octahedral unit (highlighted in purple) at the edge of the crack from the "bright" and "dim" atomic columns can rotate 90º and occupy an interstitial site in the first step and then reach its final site after another 90º rotation in the second step. Axes of rotation are indicated. Each interstitial site can be characterized by the distances from the nearest neighboring atomic column (highlighted by red dashed line). (c) HAADF-STEM image of a partially bridged crack showing presence of Ti partial atomic columns at different interstitial sites (indicated by arrows). (d) Statistical analysis of the locations of Ti partial atomic columns in the interstitial sites from HAADF images obtained from many "bright" and "dim" rows.

To further examine this "2-step" octahedral rolling model, we conducted fine-grained image analysis of electron beam irradiated areas near the cracks. Based on this rolling model, we should see many Ti atoms at interstitial sites of the rutile $TiO_2$ crystal in these regions, as the rolling octahedra pass through interstitial sites in the process of extending the crystal (as shown in Figs. 5(b), also in Fig. S8 and Video S2 in SI). This matches what we observe in atomic-resolution

HAADF-STEM images obtained from those regions (Figs. 5(c) and Fig. S9). It should be noted that, when viewed along the ⟨110⟩ direction, there are three unique and identifiable interstitial sites in rutile TiO₂ crystal where the rolling octahedra should be observed after the first rolling step: two sites between "bright" columns 1.5 and 2.0 Å away from the nearest Ti-O plane, one site between "dim" columns at 1.3 Å away from the nearest Ti-O plane. In these HAADF-STEM images of bridging regions, we identify the presence of partial Ti atomic columns in all three interstitial sites (Figs. 5(d)).

To further evaluate the radiolysis-induced restructuring process, additional experiments with different electron beam energies were performed using similar set-up (dose rate, beam convergence angle, etc.). A set of STEM experiments were conducted at $E_0$ = 80, 200, and 300 keV beam energies, using three different cracks on a single FIB-cut TEM specimen. While results show that bridging and filling occurs in all conditions, it is faster at lower beam energies (see Fig. S10 and Video S3 in SI), exhibiting the characteristic beam energy dependence of radiolysis. In a rutile TiO₂ crystal, every Ti atom is bonded to six oxygens with a bond dissociation energy of 3.3 eV/bond (see Tables S1 and S2 in SI), where four bonds have a length of 1.94 Å and the other two (opposite to each other) are slightly longer at 1.99 Å, making them more susceptible to bond breaking [16]. Since the exciton energy in rutile TiO₂ is approximately 3.0 eV [31,32] with a very long exciton lifetime of $\tau_{ex}$ = 16 ns [33], the formation of one or two excitons at a time readily enables radiolytic rolling of a single octahedron. Rolling displacements of edge and corner octahedral units are expected to be more frequent than those of surface units, as they only require breakage of 1-3 bonds, depending on whether they will carry oxygen with them or not (see Fig. S11).

The efficiency of radiolysis in rutile TiO₂ can be determined from analyzing crack-filling image series at each of the different beam energies. Using the cross-section for radiolysis [2]:

$$\sigma_r(E_0) = 8\pi a_0^2 \times \left(Z \frac{U_R}{m_0 c^2}\right) \left(\frac{U_R}{E_{th}\beta^2}\right) \times \zeta, \quad (1)$$

where $a_0$ is the Bohr radius, $Z$ is the atomic number of a moving unit, $U_R$ is Rydberg energy, $m_0$ is the rest mass of electron, $c$ is the speed of light, $E_{th}$ is the threshold energy that must be transferred to produce a movement, and $\beta = \frac{v}{c} = \sqrt{1-\left(1+\frac{E_0}{m_0c^2}\right)^{-2}}$ and fitting to the data, the efficiency of a single step rolling in the "2-step" octahedral rolling model is estimated to be $\zeta \simeq 6 \times 10^{-6}$ (see Fig. S10). This efficiency value is slightly lower than that determined for silicates ($\zeta \sim 10^{-4}$) [2], which is expected given that the octahedral unit in rutile $TiO_2$ is larger than the tetrahedral unit in silica. Since the cross-section for radiolysis is inversely proportional to the incident beam energy, the speed of the octahedral motion can be accelerated by order of magnitude using a lower-energy electron beam (~10 keV). Based on information about crystal structure, excitonic energy and lifetime, band gap, and atomic bond dissociation energies (summarized in Table S2 in SI), other oxides – anatase $TiO_2$, rutile $SnO_2$ and $GeO_2$, even α-quartz $SiO_2$ and $Al_2O_3$ – have all the essential ingredients to exhibit similar radiolysis-driven restructuring via rolling building blocks. However, carefully designed experiments are needed to confirm these predictions.

In conclusion, analysis of high-resolution HAADF-STEM images combined with core-level EELS spectra shows that high-energy (80-300 keV) electron beams at high doses ($\gtrsim 10^7$ e/nm$^2$) can radiolytically restructure a crystalline material instead of amorphizing it. Such unusual constructive radiolysis-driven restructuring was observed in rutile $TiO_2$. The unique sample geometry, wherein nanometer-scale cracks with atomically sharp edges are aligned with the STEM beam, allowed atomic-resolution visualization of this radiolysis-driven restructuring of the surrounding rutile $TiO_2$ crystal. We propose that this restructuring is the result of many "2-step rolling" movements of the octahedral units located at the exposed corners, edges, and surfaces of the crack. Channeling radiolysis as a constructive force sheds new light on atomistic mechanisms that drive radiolytic structural modifications in insulating materials. We predict that similar radiolysis-driven restructuring should occur in other oxides with crystal structures consisting of octahedral or tetrahedral building blocks, as long as they satisfy the requirements for significant radiolysis. These observations also point to new possibilities for using an electron beam to treat sharp cracks in brittle ceramics, improve the quality of wide bandgap thin films during the growth, and engineer novel nanostructures.

This work was supported primarily by the National Science Foundation through the University of Minnesota MRSEC under Award No. DMR-2011401. Parts of this work were carried out at the UMN Characterization Facility, supported in part by the NSF through the UMN MRSEC program. The authors acknowledge the Minnesota Supercomputing Institute (MSI) at the University of Minnesota for providing computational resources. Film growth was supported by the U.S. Department of Energy (DOE) through Grant No. DE-SC0020211. We wish to thank Dr. Michael Odlyzko for help with STEM experiments and providing feedback.13

# References

[1] L. Reimer and H. Kohl, *Transmission Electron Microscopy: Physics of Image Formation* (Springer, Berlin, 2008), 5th edn.
[2] L. W. Hobbs, Introduction to Analytical Electron Microscopy, edited by J. J. Hren, J. I. Goldstein, and D. C. Joy (Scanning Microscopy International, Chicago, 1979), p. 437; Scanning Microscopy Supplement 4, edited by J. Schou, P. Kruit, and D. E. Newbury (Scanning Microscopy International, Chicago, 1990), p. 171.
[3] O. Ugurlu, J. Haus, A. A. Gunawan, M. G. Thomas, S. Maheshwari, M. Tsapatsis, and K. A. Mkhoyan, Phys. Rev. B **83,** 113408 (2011).
[4] D. L. Medlin, L. E. Thomas, and D. G. Howitt, Ultramicroscopy **29**, 228 (1989).
[5] K. A. Mkhoyan and J. Silcox, Appl. Phys. Lett. **82**, 859 (2003).
[6] R. F. Egerton, P. Li, and M. Malac, Micron **35**, 399 (2004).
[7] M. N. Kabler and R. T. Williams, Phys. Rev. B **18**, 1948 (1978).
[8] Y. Y. Zhou, H. Sternlicht, and N. P. Padture, Joule **3**, 641 (2019).
[9] M. R. Pascucci, J. L. Hutchison, and L. W. Hobbs, Radiat. Eff. **74**, 219 (1983).
[10] H. Inui, H. Mori, T. Sakata, and H. Fujita, Journal of Non-Crystalline Solids **116**, 1 (1990).
[11] M. M. J. Treacy and J. M. Newsam, Ultramicroscopy **23**, 411 (1987).
[12] Y. Yokota, H. Hashimoto, and T. Yamaguchi, Ultramicroscopy **54**, 207 (1994).
[13] P. Kumar, D. W. Kim, N. Rangnekar, H. Xu, E. O. Fetisov, S. Ghosh, H. Zhang, Q. Xiao, M. Shete, J. I. Siepmann, T. Dumitrica, B. McCool, M. Tsapatsis, and K. A. Mkhoyan, Nat. Mater. **19**, 443 (2020).
[14] S. Ghosh, H. Yun, P. Kumar, S. Conrad, M. Tsapatsis, and K. A. Mkhoyan, Chem. Mater. **33**, 5681 (2021).
[15] D. L. Zhang, Y. H. Zhu, L. M. Liu, X. R. Ying, C. E. Hsiung, R. Sougrat, K. Li, and Y. Han, Science **359**, 675 (2018).
[16] F. A. Grant, Rev. Mod. Phys. **31**, 646 (1959).
[17] K. M. Glassford and J. R. Chelikowsky, Phys. Rev. B **46**, 1284 (1992).
[18] S. D. Berger, J. M. Macaulay, and L. M. Brown, Philos. Mag. Lett. **56**, 179 (1987).
[19] N. Shibata, A. Goto, S. Y. Choi, T. Mizoguchi, S. D. Findlay, T. Yamamoto, and Y. Ikuhara, Science **322**, 570 (2008).
[20] W. Nunn, A. K. Manjeshwar, J. Yue, A. Rajapitamahuni, T. K. Truttmann, and B. Jalan, PNAS **118**, e2105713118 (2021).
[21] J. P. Ruf, H. Paik, N. J. Schreiber, H. P. Nair, L. Miao, J. K. Kawasaki, J. N. Nelson, B. D. Faeth, Y. Lee, B. H. Goodge, B. Pamuk, C. J. Fennie, L. F. Kourkoutis, D. G. Schlom, and K. M. Shen, Nat. Commun. **12** (2021).
[22] S. Nair, Z. Yang, D. Lee, S. Guo, J. T. Sadowski, S. Johnson, A. Saboor, Y. Li, H. Zhou, R. B. Comes, W. Jin, K. A. Mkhoyan, A. Janotti, and B. Jalan, Nat. Nanot. (accepted).
[23] R. F. Egerton, *Electron Energy-Loss Spectroscopy in the Electron Microscope* (Plenum Press, New york and London, 2$^{nd}$ edn.
[24] L. D. Finkelstein, E. I. Zabolotzky, M. A. Korotin, S. N. Shamin, S. M. Butorin, E. Z. Kurmaev, and J. Nordgren, X-Ray Spectrum. **31**, 414 (2002).
[25] R. Brydson, H. Sauer, W. Engel, J. M. Thomass, E. Zeitler, N. Kosugi, and H. Kuroda, J. Phys. Condens. Matter **1** (1989).
[26] E. Stoyanov, F. Langenhorst, and G. Steinle-Neumann, Am. Mineral. **92**, 577 (2007).
[27] A. Gloter, C. Ewels, P. Umek, D. Arcon, and C. Colliex, Phys. Rev. B **80,** 035413 (2009).

[28] T. Tamura, S. Tanaka, and M. Kohyama, Phys. Rev. B **85,** 205210 (2012).
[29] R. Ameis, S. Kremer, and D. Reinen, Inorg. Chem. **24**, 2751 (1985).
[30] M. K. Tian, M. Mahjouri-Samani, G. Eres, R. Sachan, M. Yoon, M. F. Chisholm, K. Wang, A. A. Puretzky, C. M. Rouleau, D. B. Geohegan, and G. Duscher, ACS Nano **9**, 10482 (2015).
[31] A. Amtout and R. Leonelli, Solid State Commun. **84**, 349 (1992).
[32] L. Kernazhitsky, V. Shymanovska, T. Gavrilko, V. Naumov, L. Fedorenko, V. Kshnyakin, and J. Baran, J. Lumin. **146**, 199 (2014).
[33] Y. Yamada and Y. Kanemitsu, Appl. Phys. Lett. **101** (2012).